\documentclass[letterpaper,11pt,review]{article}  
\usepackage{osajnl2} 
\usepackage{amsmath}
\usepackage{bm}
\usepackage[pdftex,colorlinks=true,bookmarks=false,citecolor=blue,filecolor=blue,urlcolor=blue,pdfstartview=FitH]{hyperref}

\begin{document}

\title{The detection of low-energy Quantum Gravity fluctuations with entangled states.}

\author{Fabrizio Tamburini}

\address{Dept. of Astronomy University of Padova, Vicolo
dell'Osservatorio 3, IT-35122 Padova, Italy.}

\begin{abstract}
We propose a thought experiment to detect low-energy Quantum Gravity phenomena using  Quantum Optical Information Technologies.
Gravitational field perturbations, such as gravitational waves and quantum gravity fluctuations, decohere the entangled photon pairs, revealing the presence of gravitational field fluctuations including those more speculative sources such as compact extra dimensions and the sub-millimetric hypothetical low-energy quantum gravity phenomena and then set a limit for the decoherence of photon bunches and entangled pairs in space detectable with the current astronomical space technology.
\end {abstract}

 
\maketitle

\section{Introduction}

Quantum Optical Information Technologies find several applications in quantum information, computation science, quantum cryptography, quantum radiometry, involving also astronomy and biology.
Those techniques provide a stunning  practical application of Einstein-Podolsky-Rosen (EPR) correlations, entangled states and Bell's inequalities and also the tools for the creation of sources to produce single photons either on demand (photon guns) or heralded. Heralded photons, produced during a parametric downconversion process, take this name because they are ``labelled'' by the detection and counting of coincidences of their correlated or entangled twins. For a better insight, see Ref. \cite{cas}.

In this paper we propose the use of Quantum Optical Information Technologies, or, more precisely, of the ideal gravitational wave (GW) detectors based on entangled states \cite{tam00,tam08} to detect the shadowy signature of the emergence of low-energy quantum gravity (LEQG) fluctuations. 
In those setups GWs may reveal their presence behaving as shadow eavesdroppers that distort the quantum encryption key statistics away from a pure white noise after the analysis of the emerging color distortions in the key. A different approach is suggested in In Ref. \cite{tam09}, where, the presence of a gravitational wave causes a relative rotation of the reference frames associated to the double-slit and to the test polarizer of a Walborn's quantum eraser. The GW is revealed by the detection of heralded photons in the dark fringes of the recovered interference pattern by the quantum eraser. The detection of GWs still remains one of the most important, challenging and outstanding test of Einstein's General Relativity and the subject of current and next-generation experiments such as LIGO, VIRGO, the Einstein Telescope and LISA \cite{ligo1,ligo2,ligo3,geo1,geo2,virgo1,virgo2,virgo3,lisa1,lisa2,et}. GWs are wave-like solutions of Einstein's General Relativity. Up to now no traces of GWs have been observed with the current detectors. For a brief introduction see Ref. \cite{sw72}.

Quantum Gravity can be at all effects considered the ÒHoly GrailÓ of modern physics. 
Strings and other Quantum Gravity (QG) theories, such as Supergravity or Brane-World scenarios, suggest that quantum gravity may manifest at energies much below the Planck scale and could even produce macroscopically observable effects. One example is the expected production of mini-Black Holes at the Large Hadron Collider, which would imply the presence of QG fluctuations also at energies on the order of the 10TeV.

LEQG fluctuations are hypothetical phenomenological aspects that derive from String theory and other alternative formulations of quantum gravity. The possibility of detecting LEQG fluctuations with current gravitational wave detectors has been recently discussed in Ref. \cite{hog08}, but the result remains still controversial \cite{smo09}.
In Refs. \cite{amecam,ford} was instead suggested to test the effects of the quantum properties of space time with the propagation of massless particles such as photons. Those fluctuations might be revealed by detecting the loss of phase coherence in the light that progressively blurs the images of distant sources like supernovae and quasars \cite{ame94,ame00,ngv94,ng03,rtg03}, but, up to now, without any clear success \cite{str,tam09b} indicating that the energy fluctuations could occur at spatial scales much below those hypothesized by the sub--millimetric models. 

In the weak field limit, any gravitational perturbation, including a gravitational wave, can be described in terms of a small perturbation $h_{ij}$ ($|h|<<1$) occurring in a Minkowsky spacetime support, and the metric tensor becomes $g_{ij}= g^{(0)}_{ij}+h_{ij}$, where $g^{(0)}_{ij}$ is the metric tensor of the unperturbed Minkowsky spacetime background. 

We now present some key considerations for an ideal experiment in which gravitational field fluctuations may be revealed only through the cumulating decoherence that affects the quantum states.  Here we do not consider, for the sake of simplicity, the effect of instrumental noise. The reader, interested in this topic may find a deeper insight in Refs. \cite{tam00,tam08,tam09}.
During this work we use the so--called natural units, by assuming $c=\hbar=G=1$.

\section{Gravitational field perturbations and entangled states}

The detection of a general gravitational field perturbation, including those generated by LEQG, can proceed in a similar way as discussed in \cite{tam00,tam08}, where the authors analyzed the problem of building of a quantum cryptographic key in a curved spacetime. For this reason, one must first consider the general case where relativistic effects either due to random fluctuations of spacetime or to the problem of simultaneity might affect the structure of a quantum key.
Moreover one has also to take in account at those small scales also the problems due to the effective quantum mechanical aspects of the photons used as basic tools to reveal the basic structure of spacetime expected from the different Quantum Gravity scenarios.

In the general scheme of an entangled states detector, two observers, with initial relative zero velocity, the sender A and the receiver B, act on shared maximally entangled pairs of photons generated by a parametric down-conversion process
\begin{equation}
|\Psi \rangle =\frac 1{\sqrt{2}}[|1\rangle _A|2\rangle_B +e^{i\phi
} |2\rangle_A|1\rangle _B]
\end{equation}
where $|1\rangle_{A,B}$ and $|2\rangle_{A,B}$ build the basis of eigenvectors for the singlet state, and are measured in the respective local galilean reference frames of the two experimenters. The parameter $\phi$ is the phase factor of the entangled state. 

The main limits for this thought experiment and also for the building of a quantum key in a curved spacetime come directly from the wave nature of the photon itself and from the presence of the time windows that are needed to the experimenters to consider as undistinguishable the two entangled photons in each pair \cite{zeibook}. The time of arrival for each single photon is thus determined within a time indetermination interval $\Delta W$ that cannot be reduced to a quantity smaller than that of the photon coherence time itself.
Similarly to what is usually assumed for a Quantum Key Distribution scheme, both the observers build independently a record of data labelled with the time of arrival of each photon detection, and also labeling any detected quanta with the measured quantum state, either $|1\rangle$ or $|2\rangle$. If the states of a simultaneous event detection coincide, ``1'' is written, otherwise ``0''. 
After running a cycle of measurements, A and B extract the set of coincident detections from their strings of data and generate a string of ``0'' 's and ``1'' 's, whose distribution follows, in an ideal experiment performed in flat Minkowsky spacetime, that of a pure markovian process with no memory in time \cite{tam00,tam08}.

In the weak field limit, any perturbation can be Fourier-decomposed and described in the so-called transverse-traceless formulation ({\it t.t.}) like for gravitational waves. The metric perturbation becomes
\begin{eqnarray}
h_{ab}(t)=\int_{-\infty}^{\infty} df \int d \hat\Omega e^{-2\pi
ft} [\hat h_+(f,\hat\Omega)e^+_{ab}(\hat\Omega)
+ \hat h_{\times}(f,\hat\Omega)e^{\times}_{ab}(\hat\Omega) ]
\end{eqnarray}
where $f$ is the frequency, $\hat h_{+,\times}(-f,\hat\Omega)=\hat h_{+,\times}^{*}(f,\hat\Omega$) represent the unit vector of the wave propagation, $d\hat\Omega = d \cos \theta d \phi$ is the solid angle and $e^{+,\times}_{ab}(\hat\Omega)$ the polarization vectors ``$+$'' and ``$\times$'' \cite{maggiore}. 

Simultaneity, energy, momentum and polarization directions of photons are not preserved in General and Special Relativity \cite{kok03} and those effects must be taken in account by A and B to find the coincidences between each couple of events to build a cryptographic key. The gravitational perturbations change also both the distance between the two observers and the proper local times of A and B with the result of affecting the coincidences between the measured time of arrival of each photon or, conversely, the path that a photon has to travel inside an interferometer. 

If $\tau_A$ and $\tau_B$ are the proper times of a detection measured in the reference frames of the observer A and B, respectively, to cross-correlate the two strings both the observers must synchronize the first detection event according to General Relativity and the local time intervals, measured after the first detection event must be corrected by the factors 
\begin{equation}
\Delta \tau_{A,B} = \Delta t/\sqrt{-g_{00}(x_{A,B})}=\Delta t \sqrt{-g^{(0)}_{00}(x_{A,B})-h_{00}(x_{A,B})}
\end{equation}

If the two observers, instead, decide at a certain proper time $\tau_A$ or $\tau_B$, related to an external event $\{x\}$, to stop their measurements and compare their strings of local detections without synchronizing their photon counting data records, one then expects that the different time delays induced by a gravitational perturbation would also differently affect the length of each string of data:
length [{\bf A}] - length [{\bf B}] $\neq 0$, being 
\begin{equation}
\frac{length[{\bf A}]}{length[{\bf B}]}\sim \sqrt{\frac{g_{00}(x_B)}{g_{00}(x_A)}}.
\end{equation}

Together with the problem of simultaneity and of the differences between the proper local times, the experimenters must also consider the effects due to the presence of $\Delta W$, that unavoidably affects both the measure of the perturbation amplitude $h$ and that of the reduced spatial wavelength $\lambda^0_{gw}= \lambda_{gw}/2 \pi$. 
In fact, if one considers the example of a plane GW propagating in a direction perpendicular to the plane identified by the entangled state propagation \cite{amecam} than obtains
\begin{equation}
\delta h_t \simeq \frac { \Delta W}{2 \lambda_{gw}\left|\sin\left( \frac{r}{2 \lambda^0_{gw}}\right)\right|}.
\end{equation}

After having corrected the effects due to the lack of simultaneity between the two events, and following \cite{tam00,tam08}, the result of the cross-correlation $S(t)$ of the $N$ local measurements is a sequence of ``0'' 's and ``1'' 's that can be described by using the mathematical formalism of fractional brownian motions and fractional calculus. Without loosing in generality, when $N$ becomes very large and the discrete string can be approximated by a continuous process and use Ito stochastic calculus \cite{oek}. 
\begin{equation}
S(t)=s(t)+n(t) \sim Q(t) Wh_t + Wh_t,
\end{equation}
The quantity $s(t)$ is the perturbation in the string caused by the gravitational wave and $n(t)$ is the unperturbed markovian process. $Wh(t)$, which is a white noise process related to the differential of the standard brownian motion $dB_t=Wh_tdt$. Finally, the quantity $Q(x,t)$ is the local perturbation. Of course without GWs/LEQG fluctuations the distribution of voids and detections is a pure white noise, with zero average.
In the presence of a gravitational perturbation, the difference between detections and non detections is expected to be different from zero 
\begin{equation}
\left|N_{[1]}-N_{[0]}\right|= \frac 12 \int_{t_0}^{T}[1+Q(x,t)]dB_t
\end{equation}
and the stochastic differential term $dB_t=Wh^{\prime}_tdt$ must be expressed in terms of the proper local time of each the observers. The observer A, for example, obtains
\begin{eqnarray*}
\left|N_{[1]}-N_{[0]}\right|&=& \frac 12 \int_{\tau_{A,0}}^{T_A} Wh'_{\tau_A}d
\tau _A + \frac 12 \int_{\tau_{B,0}}^{T_B}  Q(x,\tau_B)
Wh'_{\tau_B}d \tau_B
\\
&=& \frac 12 \int_{\tau_{A,0}}^{T_A} d\tau_AWh_{\tau,A}
\left[1+\sqrt{\frac{g_{00}(x_A)}{g_{00}(x_B)}} Q(x,\tau_A)\right]
\end{eqnarray*}
that in the weak field approximation becomes
\begin{equation*}
\left|N_{[1]}-N_{[0]}\right| \simeq  \frac 12 \int_{\tau_{A,0}}^{T_A}
dB_{\tau_A}\left\{1+\sqrt{\frac{g_{00}(x_A)}{g_{00}(x_B)}}\left[1- \frac {
\Delta W}{4 \lambda_{gw}\left|\sin\left(\frac{r}{2
\lambda^0_{gw}}\right)\right|}\right]\right\}
\end{equation*}
showing that any gravitational field perturbation can be put in evidence through the discoloration of the white noise of the coincidence detections. Here $r$ s the distance between the two observers and $\lambda_{gw}$ the wavelength of the plane GW. 

\section{Low-energy Quantum Gravity effects in a Minkowsky spacetime}

We now proceed to analyze how to detect LEQG effects with entangled photon states.
LEQG lightcone fluctuation effects in a flat spacetime may give a way to detect the nontrivial topology with compactified spatial dimensions. Many different approaches to QG describe different models of ``fuzziness'' of spacetime expressed as low energy deformed dispersion relationships of the spatial distance $r$ between two events for a given energy value $E$, namely, $
\sigma_r \sim L_{min}$.
The amplitude spectral density associated to the displacement induced by spacetime fuzziness is given by
\begin{equation}
S_{min}(f) \sim \frac{L_{min}}{\sqrt{f}}
\end{equation}
and the relationship between the Root Mean Square (RMS) value $\sigma$ and $S(f)$ is
\begin{equation}
\sigma^2 = \int^{f_{max}}_{1/T_{obs}}\left[S(f)\right]^2df.
\end{equation}

In the more general case, the relativistic formulation of energy-momentum relationships result modified by the presence of low-energy quantum gravity effects 
\begin{equation}
{\bf p}^2 \simeq E^2\left[1+ \xi \left( \frac{E}{E_{QG}}\right)^{\alpha}\right]
\end{equation}
where $\alpha$ and $\xi$ are the two parameters that describe at the first order the model of quantum spacetime fluctuations. $E_{QG}$ is the scale of QG, one expects that those quantum effects become observable.
The three main classes of models of QG are obtained by varying $\xi$ and $\alpha$. 
In the Random walk model, $\alpha=1/2$, the random perturbations of space-time add incoherently with a square root dependence\cite{ncv03} (or with some other more general stochastic process \cite{ame00,ngv94}). 
The value $\alpha=2/3$, instead, describes spacetime foams consistent with WheelerÕs and Beckenstein-Hawking's holographic principle\cite{tho93,sus95,mal98} and with Black Hole entropy. 
Finally, $\alpha=1$ describes the standard model of Loop QG with the Planck time as characteristic 
time of the fluctuations \cite{ngv00,nlo01,rov04}.

If $E_{QG} \ll E_{Planck} \sim 10^{16}~Gev$, LEQG effects should affect the propagation of photons, and consequently their paths in the interferometer. The dispersion relationship on photon arrival times becomes
\begin{equation}
\sigma_D \sim \sqrt{\left( \frac{\alpha+\alpha^2}{2}\right){\left( \frac{E_{typ}}{E_{QG}}\right)}%
^{\alpha -1}\left(\frac{ T_{obs}}{E_{QG}}\right)},
\end{equation}
where $E_{typ}$ is an energy scale that characterizes the physical context
that we are considering, and we suppose that the spectrum of the noise
produced by the device is considered negligible in the ideal case.

In this case the string distribution is a simple white noise with a spectral density
amplitude dependent on the frequency $S(f) \sim f^{-1}$. The (RMS) is a
function of the observation time $\sqrt{T_{obs}}.$ When the deviation $\sigma_D \sim T^{1/3},$ the spectral density becomes $S(f) \sim f^{-5/6}$
thus affecting the value of the correlation parameter $\alpha$.

Let $h_{\mu \nu}$ be a linearized three-dimensional metric perturbation. The three-dimensional interval $\sigma$ between the two observers can be expanded at the first order as
$\sigma_0 + \sigma_1 + O(h_{\mu \nu}^2)$ that is, following Ref. \cite{ford}, 
$\sigma = \frac 12 [(r+\Delta t)^2 - r^2]\approx r\Delta t$.
The influence of the compactification of extra dimensions on the propagation of the light is given as a function of the paths traveled by the photons along their geodesics $r=a-b$ and of a dimensionless parameter $\epsilon=r/L$, which is $\Delta t \approx \sqrt{\frac{2\zeta(3)G_4}{\pi}}\epsilon \approx t_{pl} \epsilon$, 
where $t_{pl}$ is the Planck time, $\zeta(3)$ is Riemann's zeta function and $G_4=1$ is the Newton's constant in 4 dimensions, in natural units.
In the rescaled time $t^{\prime}$ the string is again represented by an uncorrelated Markovian process with null correlation if $\alpha = 1$.

\subsection{Signal detection via stochastic noise analysis}

As already said, the effect caused by gravitational fluctuations is very small, and the only way the experimenter has to give evidence to them, either stochastic or deterministic, such as a plane wave, is to accumulate progressively in a string of ``0'' 's and ``1'' 's the result of their joint measurements performed on the shared quantum states. 
Instead of looking inside each single q-bit and calculating for each element in the string the probability of a q-bit mismatch, one should consider to analyze instead the color of the noise. Each fluctuation will modify the shape of the random key away from the pure white noise. For more details about the detection of stochastic background and the problem of noise in the experiment see Ref. \cite{tam08}.
To give an example, the string of data record can be modeled as the superposition between a pure markovian process and a stochastic background noise or a deterministic perturbation. Off-power terms given by the cross-correlation of the two strings of A and B would reveal a deterministic component (see Fig.1 in Ref. \cite{tam08}). 

If the perturbing term is a stochastic noise, instead, one should consider to perform a correlation analysis of the strings.
A practical way to handle time series made with discrete data and characterize their statistical behavior away from the white noise is Hurst's analysis and fractal geometry. The fractal dimension of the noise, namely the color is determined by a time series analysis such as Hurst's analysis \cite{hur51,hur65}.
Mandelbrot and van Ness (1968) and Mandelbrot and Wallis (1969) linked this method to a particular class of self--similar random processes, called Fractional Brownian Motions (FBMs) \cite{low99,gao03}. 
From this point of view, the behavior caused by the spacetime fluctuations on both the time series, obtained from the markovian process generated by the interferometer's setup, can be modeled by a discrete version of a FBM with index $\beta \in (0,1],$
\begin{eqnarray*}
B_{\beta /2}(t) - B_{\beta /2}(t-1)
&= &\frac{n^{-\beta /2}}{\Gamma(\beta /2 +0.5)} \left\{ \sum_{i=1}^n
i^{\beta /2 -0.5}\xi_{[1+n(M+t)-i]}+ \right. \\
&+& \left. \sum_{i=1}^{n(M-1)}((n+i)^{\beta /2 - 0.5}- i^{\beta /2
-0.5}\xi_{[1+n(M-1+t)-i]} \right\}
\end{eqnarray*}
where $\xi_i$ is the set of gaussian random variables with variance $1$ and mean $0$, $M$ is the step, $n$ the number of records, and $\beta /2$ the correlation parameter. Depending on the value of $H$ found, the behavior of the random fluctuations can be modeled by Wiener stochastic processes or, more specifically, by Fractional Brownian Motions (FBM).
In dynamical systems, $H$ characterizes the stochastic memory in time of the process. Hurst exponents $H> 1/2$ indicate the persistence in time of the stochastic process, whereas exponents $H< 1/2$ indicate anti-persistence, i.e., past trends would tend to reverse in the future.
An exponent $H=1/2$ would represent random uncorrelated behaviors with no stochastic memory in time, described by a classical Brownian motion, related to the pure white noise \cite{fed88}. 
Since FBMs possess self-similarity, they can also be also easily studied with the wavelet analysis \cite{chui92,gil90,sim98}. 
By definition, the power spectrum of the stochastic noise for LEQG effects is given by the parameter $\alpha$. The power spectrum of FBM is defined by the parameter $\beta$. The power spectrum of a FBM obeys a power law
$S(\omega) \sim c/\omega^\alpha$ for large frequencies $\omega$ of the Fourier transform of a FBM \cite{fed88}. The relationship between the two parameters $\alpha$ and $\beta$ is quite straightforward, by directly replacing $\alpha \rightarrow \beta/2$, can put in evidence the noise generated by a specific model of Quantum Gravity and determine the shape of spacetime foam.

We have seen that the time indetermination in the photon arrival time is unavoidably affected by the photon coherence length, that would require shorter and shorter wavelengths to reduce $\Delta W$.
For this reason, instead of considering the variations of very short time intervals caused by gravitational field fluctuations either to label the result of each measurement on quantum states or to directly estimate the gravitational field fluctuation, the experimenter can instead use a Michelson interferometer and try to detect the GW by measuring the deformation in the interferometer's arms due to GWs or LEQG effects.
Here photons are entangled in a superposition of states $|s\rangle _p$ and $|l\rangle _p$ characterized by having chosen either the short (\textit{s}) or long (\textit{l}) paths in the interferometer \cite{tittel98}. 
\begin{equation}
|\Psi \rangle =\frac 1{\sqrt{2}}\left[|s\rangle _A|s\rangle _B+e^{i\phi}|l\rangle _A|l\rangle _B\right]
\end{equation}
See also \cite{zeibook} for a deeper insight.

A coincidence detection is achieved when both photons pass through the short arm $|s\rangle $ or the long one $|l\rangle $.
Entangled photons distribute along three peaks, that correspond to a superposition of short/short, short/long and long/long pahts, respectively. By measuring the number of photons present in each of the paths, one can easily build a purely markovian process associated to the interferometer's output.
The time-delay effect of a gravitational wave $\delta t^2\simeq \frac 1{1+h_{00}}
\{h_{0\alpha }h_{0\beta }-h_{\alpha \beta }(1+h_{00})\}$
results on a stretching of one or both the interferometer's arms of a quantity 
$\delta l^2\simeq \left(\frac{h_{0\alpha }h_{0\beta }}{1+h_{00}}-h_{\alpha \beta}\right)dx^\alpha dx^\beta$ that affect the basis $\{|s\rangle ,|l\rangle \}$ in which photons are entangled, moving away photons from their initial location, with the result of coloring the stochastic markovian process.
The difference between the number of coincidences along the short and long paths, $|s\rangle|s\rangle$ and $|l\rangle|l\rangle$, respectively, distribute as a FBM and reveals the quantum properties of the spacetime as already discussed.

\section{Conclusions}
We proposed a thought experiment to detect low-energy Quantum Gravity fluctuations with Quantum Optical Information Technologies. Two observers located in two different regions of the spacetime perform joint measurements on the shared entangled states and build a string of 0 and 1 by measuring the detection coincidences. In a Minkowsky flat spacetime the sequence of symbols in the string distribute like a pure white noise. Gravitational field fluctuations de-synchronize the set of joint measurements between the two observers  and change the distribution of ``0'' 's and ``1'' 's in the string with the result of coloring the noise. 
Based on the results obtained from Refs. \cite{str,tam09b}, we can easily infer that, if the sensitivity needed to detect those effects must be extremely high, the quantum states will be preserved for billions of light years within the errors dictated by the current technology present in the Hubble Space Telescope. Quasars distant more than 8 billions of light years did not show any blurring caused by LEQG effects. From these results one obtains an upper limit to the phase factor imposed by QG fluctuations on cosmological distances which is on the order of $\delta \phi \sim 2.15\times 10^{-13}$ if Alice is on the Earth and Bob on Mars, being the upper limit found for $\alpha=0.67$. Fluctuations at the Planck scale do not cumulate and fix a value $\delta \phi \sim 6.15\times 10^{-17}$.

\textbf{Acknowledgments}\\
The author gratefully acknowledges the financial support from the CARIPARO Foundation inside the 2006 Program of Excellence.

\end{document}